Robert Long
NYU Department of Philosophy
rl2898@nyu.edu
5 May 2020

# Fairness in machine learning: Against false positive rate equality as a measure of fairness

*Abstract:* As machine learning informs increasingly consequential decisions, different metrics have been proposed for measuring algorithmic bias or unfairness. Two popular "fairness measures" are calibration and equality of false positive rate. Each measure seems intuitively important, but notably, it is usually impossible to satisfy both measures. For this reason, a large literature in machine learning speaks of a "fairness tradeoff" between these two measures. This framing assumes that both measures are, in fact, capturing something important. To date, philosophers have not examined this crucial assumption, and examined to what extent each measure actually tracks a normatively important property. This makes this inevitable statistical conflict—between calibration and false positive rate equality—an important topic for ethics. In this paper, I give an ethical framework for thinking about these measures and argue that, contrary to initial appearances, false positive rate equality does not track anything about fairness, and thus sets an incoherent standard for evaluating the fairness of algorithms.

## 1. *Introduction*

Imagine that you have been arrested for stealing a bike. While you are in jail awaiting a hearing, the "pretrial services staff" pulls information about you: your age, and any past convictions. This information is fed to an algorithm called COMPAS, which outputs a score that reflects how likely you are to be rearrested in the near future, from 1 to 10. [1] Judges use this score in deciding whether you will be released on bail. Jurisdictions that use these scores, known as risk assessments, hope that they can make the pre-trial decision process more efficient, more accurate, and more fair.[2]

In a widely cited, Pulitzer-nominated article, ProPublica called the COMPAS algorithm unfair and "biased against blacks". ProPublica reported that, after auditing COMPAS predictions in Broward County, FL, they had found that:

> The formula was particularly likely to falsely flag black defendants as future criminals, wrongly labeling them this way at almost twice the rate as white defendants. White defendants were mislabeled as low risk more often than black defendants.[3]

ProPublica called COMPAS biased against black defendants because of two metrics:

[1] COMPAS—Correctional Offender Management Profiling for Alternative Sanctions—is sold by a company called Equivant, which was called Northpointe at the time of the ProPublica story. This software makes use of 137 features, including responses to a questionnaire. But according to Equivant, in predicting recidivism, the algorithm makes use of only 6 features for pretrial detention. See below for more discussion of the features that COMPAS uses.

[2] "Pretrial" is a misleading term since the majority of defendants never go to trial – many plead guilty, in part because of the hardships of pretrial detention. See Stevenson and Mayson (2017).

[3] Angwin et al. (2016, 23)

**Prediction Fails Differently for Black Defendants**

| | WHITE | AFRICAN AMERICAN |
|---|---|---|
| Labeled Higher Risk, But Didn't Re-Offend | 23.5% | 44.9% |
| Labeled Lower Risk, Yet Did Re-Offend | 47.7% | 28.0% |

*Overall, Northpointe's assessment tool correctly predicts recidivism 61 percent of the time. But blacks are almost twice as likely as whites to be labeled a higher risk but not actually re-offend. It makes the opposite mistake among whites: They are much more likely than blacks to be labeled lower risk but go on to commit other crimes.* (Source: ProPublica analysis of data from Broward County, Fla.)

COMPAS does not use defendant race as an input. But these lopsided numbers look like strong evidence that COMPAS predictions are skewed by race in some problematic way. As a result, many discussions of "algorithmic bias" use COMPAS as a cautionary tale.

At the same time, outside observers and the makers of COMPAS pointed out that COMPAS does have a prima facie desirable property: COMPAS scores are calibrated between races. This means that a given COMPAS score corresponds to the same frequency of rearrest, regardless of the race of the defendant. For example, defendants who have a COMPAS score of 8 are equally likely to be rearrested, whether they are black or white. As COMPAS's makers put it, "COMPAS predicts recidivism in a very similar way for both groups of defendants....a given COMPAS score translates into roughly the same chance of recidivism, regardless of race".[4]

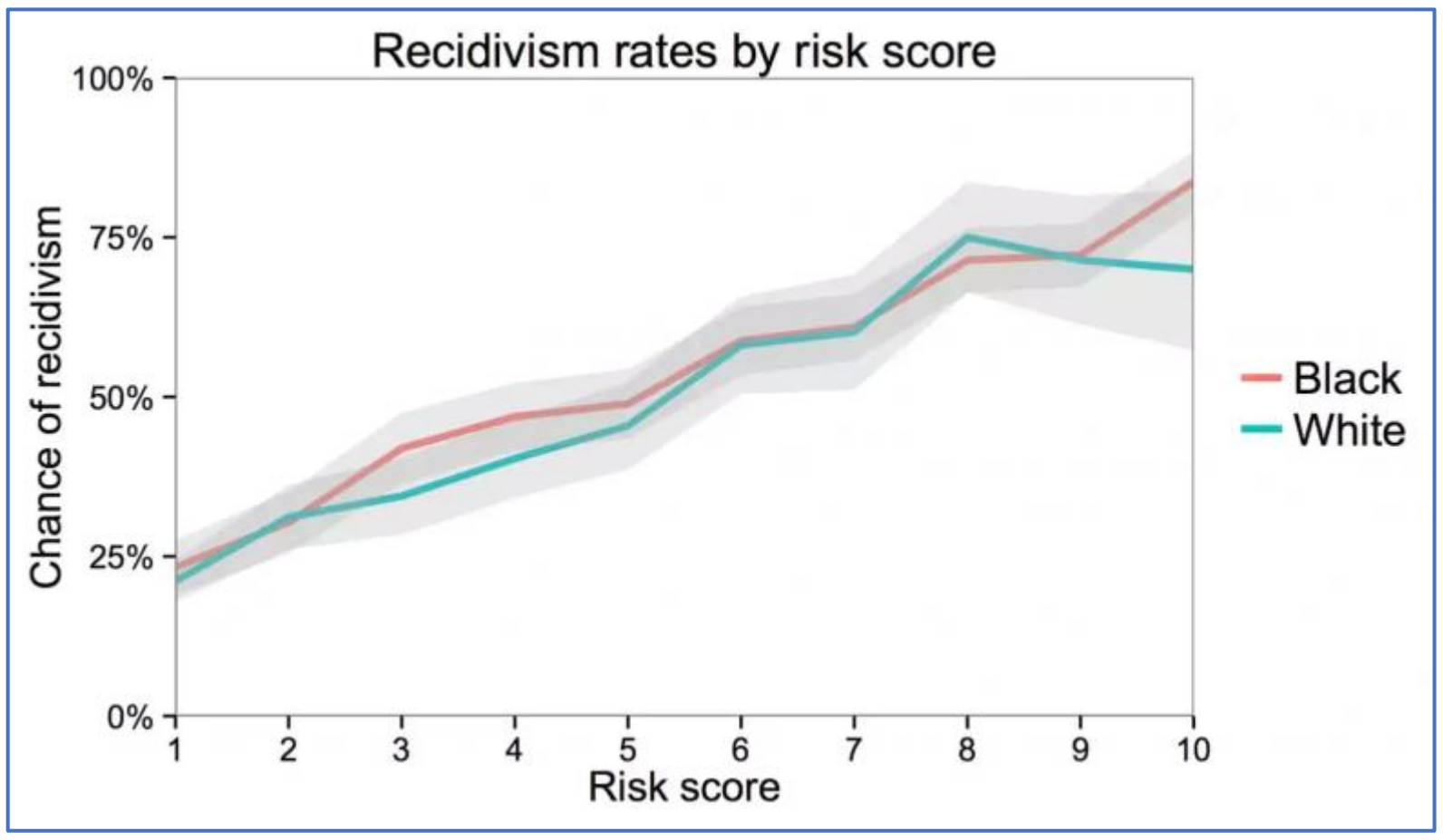

[5]

How is calibration compatible with the skewed numbers that ProPublica reported? ProPublica took the 1-10 COMPAS scores and binned them into two coarse-grained categories: "low risk" (1-4) and "high risk" (5-10).[6] Using these two categories, they defined a *false positive* prediction as a "high risk" defendant who was not rearrested, and a *false negative* prediction as a "high risk" defendant who was not rearrested. The figures reported above use those prediction categories:

[4] Flores et al. (2016, 44).
[5] Chart from Corbett-Davies et al. (2016)
[6] COMPAS does not employ these categories.

| | "Low Risk" | "High Risk" |
|---|---|---|
| **Not rearrested** | true negative | false positive |
| **Rearrested** | false negative | true positive |

ProPublica focused on two metrics defined in terms of these categories:

False positive rate (FPR): The proportion of not-rearrested defendants who are flagged as "high risk", that is, who are *false positive* predictions.

$$\text{FPR} = \frac{\text{"}high\ risk\text{"}\ \&\ not\ rearrested}{not\ rearrested} = \frac{false\ positives}{not\ rearrested}$$

False negative rate (FNR): The proportion of rearrested defendants who are flagged as "low risk", that is, who are *false negative* predictions.

$$\text{FNR} = \frac{\text{"}low\ risk\text{"}\ \&\ rearrested}{rearrested} = \frac{false\ negatives}{not\ rearrested}$$

*The normative puzzle*

These proportions are unequal for white defendants and black defendants. This seems unfair. Yet the scores are calibrated: this also seems like something that fairness demands. Thus, intuitively, each of these metrics—calibration, and the false positive and false negative rates—seems to capture something about fairness. And indeed, the debate about COMPAS involves different commentators invoking their own favored *measures* of fairness. For example, ProPublica assumes that unequal false positive and false negative rates are a mark of unfairness; on the other hand, the makers of COMPAS protest that calibration means their tool is fair. Fundamentally, people are disagreeing about the demands of fairness: should our tools be calibrated? Should we equalize false positive and false negative rates? This debate raises a more general ethical question: how should we assess the fairness or unfairness of algorithmic prediction and decision-making procedures?

One natural position is pluralism: both calibration and equal false positive and negative rates are desiderata of fairness. Such a position would merely forestall the difficult questions, however. That's because unequal error rates are not merely *compatible* with calibration: they are almost always a *necessary consequence* of calibration. In general, a calibrated predictive tool *must* produce unequal false positive and false negative rates, whenever it is applied to groups that have different base rates of the property being predicted. In the ProPublica case, that property is rearrest: black defendants have a higher base rate of rearrest than white defendants in Broward County, FL.

In such cases, you can't satisfy both measures at once. Thus, if both are necessary for fairness, then we would have a striking philosophical result: whenever groups have unequal base rates—which is almost always—then we face a moral dilemma, because we will either have the unfairness of miscalibration or the unfairness of unequal error rates. In fact, this is how ProPublica frames the issue: they write that bias is "inevitable", and that calibrated algorithms like COMPAS are "simultaneously fair and unfair".[7] And indeed, it has become common in machine learning to speak of an inevitable "fairness tradeoff" between the two measures of fairness;[8] papers claim to examine "competing notions of what it means for a probabilistic classification to be fair"[9]. Furthermore, a many papers in this literature propose methods for trading off calibration and false positive rate equality. This indicates that many people have assumed that both measures are desiderata, at least some of the time.[10]

[7] Angwin and Larson (2016)

Thus, a large literature in data ethics is built around an assumption, explicit or implicit, that both measures are, in fact, measures of fairness. To date, philosophers have not examined this crucial assumption, and asked to what extent each measure really does capture a normatively important property that deserves the name 'fairness'.[11] This makes the inevitable conflict between calibration and false positive (and negative) rate equality an important philosophical topic. For one thing, it is potentially a previously undiscovered and undiscussed moral dilemma. Furthermore, if it is indeed a dilemma or "tradeoff", then we must determine how to this tradeoff ought to be made. This will have far-reaching consequences for the use of predictive algorithms: in pre-trial detention, employment, loans, policing, and more.[12]

In this paper I argue that, contrary to initial appearances, equality of false positive and false negative rates is not directly relevant for fairness. (I focus on false positive rate inequality, rather than false negative rate inequality; analogous arguments can be made about false negative rate equality). False positive rate equality does not track anything about fairness, and sets an incoherent standard for evaluating the fairness of algorithms. This is not to say that the use of the COMPAS algorithm is fair! Indeed, moving away from the common focus on false positive rates will help us more clearly diagnose the unfairness in pre-trial predictions and decisions.[13]

First, I explain the relationship between calibration and false positive rates. Then I make the following claims. First, any predictive decision procedure subjects individuals to the risk of incorrect decisions, and a fair procedure is one that appropriately weighs the interests of individuals who are subjected to this risk. This appropriate weighting takes the form of appropriate decision thresholds. Secondly, when appropriate decision thresholds have been set, calibration is a necessary condition for procedural fairness. Third, in contrast, equal false positive (or false negative) rates are neither necessary nor sufficient for procedural fairness. An important upshot of these claims is that there is no reason to equalize false positive rate equality *per se*, and indeed doing so will not tend to make decision procedures more fair.

## *2. Understanding the central impossibility*

In this section, I outline the measures at issue and try to explain intuitively how calibration and equal false positive rates are directly at odds whenever groups have different base rates. I will use a simplified and less morally-laden example as my case study, in order to fix the relevant statistical intuitions. However, we will return to the COMPAS case and its many important complications. (Readers who are already familiar with these fairness measures may skim or skip to section 3).

---

8 Corbett-Davies et al. (2016), Corbett-Davies et al. (2017)
9 Kleinberg et al. (2016)
10 Even if people do often claim that they are leaving open the normative. For example, Berk et al. (2017, 3): "No effort is made here to translate formal definitions of fairness into philosophical or jurisprudential notions".
11 This in spite of a recent and welcome increase of philosophical treatments of topics related to fairness, ethics, and algorithms—including papers which point out that there are many important ethical and political issues beyond of these fairness measures.
12 Even sticking to the criminal justice context, we will face this question with many social groups besides race. For example, the base rate difference between men and women is far greater than any racial base rate differences.
13 For an excellent primer on various pitfalls of predictive processes, see Barocas and Selbst (2016).

*2.1. Explaining the measures*

In general, a predictive tool is a function from known features, to a predicted probability of an unknown feature. Thus, the tool outputs an estimate of a conditional probability—of an individual with known features X having unknown feature Y, P(Y = y | X = x). Often, predictors instead output a score S, which is associated with a given conditional probability.

*Case: Stride-Height predictor*

> Suppose that we must decide to who to exclude, and who to include, on a fun spelunking trip. Suppose that we don't know the heights (Y) of individuals. But we do know the lengths of their strides (X). Stride is an imperfect but reliable predictor of height. Using stride, we want to predict who is over six feet tall, or too tall for spelunking (TTFS), and thus who should be excluded from the trip. The score S assigned by this simple predictor is simply the stride length, and each stride score is associated with a given probability, or $p_{score}$, that an individual with that stride score is TTFS.

Let's suppose that stride correlates with height in the same way for men and for women. For example, a stride of 160 corresponds to the same $p_{score}$ of being TTFS, say 80%, regardless of sex. This is, stride length is a calibrated predictor of height. Suppose also that we divide everyone into "high risk" (stride at or above 160) and "low risk" (stride below 160) categories.

In general, a calibrated predictive tool like this will produce unequal false positive rates whenever two groups have different base rates of the property that is being predicted.[14]

*Central impossibility (predictions):* if a calibrated predictor is applied to two groups, then the false positive rate will be higher for the group with the higher base rate.

In this case, that base rate difference is differences in height: men are on average 5 inches taller than women, and on average their strides are about 10 inches longer.[15] Thus, the false positive rate will be higher for men.

This may seem surprising. The key is to note that calibration and false positive rate both track the number of false positives, but as shares of different things. Calibration tracks the number of false positives as a share of *predictions*: $p_{score}$ measures how likely a given *score* is to correctly pick out a tall person, and calibration means that a stride score is equally likely to do so irrespective of sex. But the false positive rate measures false positives as a share of a given *true outcome***.** Consider how the false positive rate is calculated:

$$\text{FPR} = \frac{FP}{not\ TTFS}$$

[14]As a result of its generality, the central impossibility has been discussed long before the latest discussion in machine learning ethics (to my knowledge, these discussions have been largely ignored by the recent literature). To take just one example, Hunter and Schmidt (1976) point out: "the two definitions [roughly, calibration and equal FPR] almost invariably conflict." See also Wigdor and Hartigan (1989)'s report for the National Research Council's Committee on the General Aptitude Test Battery.

[15]In the United States.

Base rate differences mean that the stride-height predictor will flag a higher share of men as "high risk"—this drives up both true positives and false positives—and that a smaller share of men are *in fact* not TTFS. So the false positive rate for men is necessarily higher.

For example, our calibrated tool might produce the following false positive rates.

$$\text{FPR women} = \frac{20\ false\ positive\ women}{100\ women\ who\ are\ not\ TTFS} = 20\%$$

$$\text{FPR men} = \frac{40 false\ positive\ men}{80\ men\ who\ are\ not\ TTFS} = 50\%$$

Since there are more positives overall (TTFS men), there are more false positives, which drives up the numerator. And the share of males who are not TTFS is lower. This drives down the denominator. So unequal false positive rates arise from the application of a calibrated tool to two populations with different base rates.

It must be emphasized that false positive rate inequality does *not* mean that a given score is more likely to yield a false positive for a man.[16] It just means that the share of false positives *as a share of not-tall men* is higher. False positive rate inequality is not a symptom of a tool *overpredicting* tallness for men relative to women—in fact, it is a result of it *not* doing so.[17]

*2.2. Decision thresholds*

Now that this intuition is clear, we can examine the normative import of these measures. When predictions inform high-stakes decisions, it is primarily these decisions themselves that matter.[18] When a predictive tool is called "unfair", this is often shorthand for whether it is liable to produce unfair outcomes as part of a decision-making process. It is this overall process—not the predictions alone—that can potentially harm people, or violate their rights, or entrench inequality.

For example, the stride-height tool informs decisions that can help or harm people. Excluding is the "correct" action for those who are over six feet tall, because the trip will be dangerous for them. But it is the "incorrect" action for those who are under six feet tall, since this unnecessarily deprives them of a good. Just as we defined false positives and false negatives for predictions, we can define them for decisions. When there are two possible actions, which can be "correct" or "incorrect" depending on the unknown predicted feature[19], we have:

| | Not exclude | Exclude |
|---|---|---|
| Not TTFS | True negatives | *False positives* |
| TTFS | *False negatives* | True positives |

16 But note that Angwin and Larson (2016)'s framework would have it that our height predictor has "been written in a way that guarantees men be inaccurately identified as TTFS more often than women".
17 In contrast, imagine a miscalibrated score: weight. This would overpredict the height of women relative to men—a man who weighs 120 pounds is likely shorter than a woman who weighs 120 pounds.
18 I say "primarily" because I don't mean to preclude that the predictions themselves can constitute a kind of wrong.
19 A considerable complication of the COMPAS decision case is that the outcome is not independent of the action, muddying the notion of the "correct" action. I don't dwell on this complication, not because it is not important, but for reasons of simplification.

A false positive decision, for instance, is a decision to exclude someone "incorrectly", when they are not in fact TTFS.

Because of the importance of decisions, in most of what follows I talk not about score calibration, but about *decision thresholds* that relate scores to decisions. (See below for how talk of calibration and talk of thresholds are interchangeable). Let's consider the simplest kind of *decision procedure*: a uniform decision threshold. For example: exclude any and all striders who have a stride of 160 or more. Applying this uniform decision threshold divides the population into two decision groups: excluded ("high risk") and not excluded ("low risk"). In such a case we recover the exact same relationship between false positive rates and calibration:

> Central impossibility (decisions): if a uniform decision threshold is applied to members of two groups, using a calibrated predictor, then the false positive rate will be higher for the group with the higher base rate.

Again, that group will be men. And again: scores do not overpredict height for males, nor are decisions being made about males in a systematically different way. In fact, the false positive rate inequality arises precisely because of calibration, a uniform decision rule, and unequal base rates. Given unequal base rates, the only way to equalize false positive rates would be to apply *differential decision thresholds:* that is, to apply a relatively lower decision threshold for women. That is, there must be some strides at which we exclude women, but include men, despite these men and women being equally likely to be TTFS.

*Case: COMPAS*

The numbers that ProPublica reported behave in exactly the same way as the stride height numbers—though of course, the properties involved are much more biased and problematic than stride and height, as I discuss later. (In particular, COMPAS purports to measure *recidivism,* but of course all it has access to is rearrest; see section 5.) The numbers reported by ProPublica arise because of two facts: COMPAS scores are calibrated, and black defendants have a higher base rate of rearrest.[20] Because of this base rate disparity, the false positive rate for black defendants is necessarily higher.

$$\text{FPR} = \frac{FP}{not\ rearrested}$$

There are more positives—both true positives and false positives—among black defendants. This drives up the numerator of the false positive rate. The share of black defendants who are not-rearrested is lower. This drives down the denominator.[21]

$$\text{FPR black defendants} = \frac{805\ false\ positives}{1795\ defendants\ not\ rearrested} = 44.9\%$$

$$\text{FPR white defendants} = \frac{349\ false\ positives}{1488\ defendants\ not\ rearrested} = 23.5\%$$

Of course, COMPAS informs a much more high-stakes decision: pre-trial detention. Temporarily bracketing doubts about what, if anything, could justify pre-trial detention, let's assume that pre-trial detention can be the "correct" action for defendants who will be rearrested if they are not detained, and is the "incorrect" action for those defendants who will not be rearrested.

[20] In Broward County, FL, the overall rearrest rate for black defendants is 51%; the overall rearrest rate for white defendants is 39%.

[21] The analogous structure holds for false negative rates being lower for black defendants.

| | Not detain | Detain |
|---|---|---|
| Not rearrested | True negatives | *False positives* |
| Rearrested | *False negatives* | True positives |

As with the stride-height case, if decision-makers apply a uniform threshold – for example, if they detain all and only defendants with scores of 8 and above, we will have the central impossibility: the false positive rate for black defendants will necessarily be higher. The only way for this procedure not to produce false positive rate inequality would be for decision-makers to apply *differential decision thresholds:* that is, to apply a relatively lower decision threshold for white defendants. That is, there must be some COMPAS scores at which we detain white defendants, but do not detain black defendants, despite these defendants being estimated as equally likely to be rearrested.

Any discussion of unequal false positive rates as a measure of fairness must begin with this fact: a demand for false positive rate equality is, necessarily, a demand for a relatively lower decision threshold for the group with the lower base rate (or a relatively higher decision threshold for the group with the higher base rate).[22] This is true, no matter the decision context. For example, consider a case I call COMPAS + benefit: if decision-makers aim to give a cash transfer to all and only those who are at a high risk of being rearrested. Achieving false positive rate equality in the COMPAS + benefit case would mandate setting a much higher threshold for distributing the benefit to black defendants!

*2.3. The central impossibility and proxy discrimination*

Before proceeding to the main claims, it's worth noting the connection between these fairness measures and related issues of proxy discrimination and statistical evidence.

In real-world domains, many people have taken false positive rate inequality to be a symptom of proxy discrimination: they believe that it can only occur if the algorithm is somehow detecting group membership via proxy attributes.[23] The stride-height predictor makes it vivid that this is not true: the predictor does not rely on any statistical correlation between being a man and being tall. It only uses only one feature, stride. Indeed, it produces false positive rate inequality precisely because it does *not* detect sex—in fact, it is *correcting* for false positive inequality which would require taking sex into account.

Similarly, the COMPAS algorithm does not take race into account. COMPAS likely[24] uses the following features: "age, sex, number of juvenile misdemeanors, number of juvenile felonies,

---

[22]What about changing not the decision thresholds, but the scores? If we hold the decision threshold fixed and change the scores—for example, by subtracting 10 from the stride scores of women—this is the same as holding the scores fixed and changing the decision threshold. Although selectively changing thresholds can always be translated into selectively changing scores, I prefer to speak in terms of changing thresholds. If scores are *transparently changed*—if decision-makers know that they have been adjusted—then it is not clear that this will change the decision procedure. The meaning of the scores changes; decision makers can update decisions accordingly.

[23]Coldewey (2018) claims that the COMPAS algorithm shows a "mysterious form of racial bias"; the BBC (2019) called COMPAS a case of algorithms discriminating based on race or race proxies.

[24]I say likely because the COMPAS algorithm is proprietary. This itself raises ethical questions about transparency, a rich topic in its own right. See, among others Jung et al. (2017), Vredenburg (ms), Creel (ms), Lipton (2016)

number of prior (nonjuvenile) crimes, crime degree, and crime charge."[25] Of course, facts about arrests (unlike facts about stride) can be deeply biased, as we will discuss. But this is a separate question from the question of whether COMPAS is using "race" proxies, which false positive rate inequality does not in itself imply.[26]

On a related note, we can ask what features one may permissibly use to make predictions. For example, one might think that it is wrong for COMPAS to use certain features, like sex and past arrests. Statistical evidence is another thorny philosophical topic in data ethics.[27] But false positive rate equality is a separable issue: we can ask whether false positive rate inequality is *itself* unfair, even in cases such as the stride-height case, when the features used are innocuous. And indeed, many people have argued or assumed that false positive rate inequality *itself* is unfair.

## 3. *Calibration, decision thresholds, and fairness*

A decision process, as we have seen, subjects individuals to risk of the harms (or rights violations) of "incorrect" decisions. The question is what statistical measures indicate that this is being done unfairly to members of a particular group.

In this section, I argue that the fairness of decision procedures is properly thought of as a question about appropriate *decision thresholds*. And I argue that, if decision thresholds have been appropriately decided, score calibration is a necessary condition for a fair procedure. That's because if scores are miscalibrated, then this is equivalent to implementing inappropriate decision thresholds, which subjects individuals to inappropriate disparate treatment.[28]

Relatedly, calibration alone is not sufficient for procedural fairness. A calibrated score can be embedded in an unfair decision process: either via inappropriate thresholds, or by a decision process that is unjustifiably inaccurate.

### *3.1. The normative nature of thresholds*

In all of the cases we have seen, a certain decision—exclude, detain, benefit—is taken under risk. Assume for idealization that the $p_{score}$ is known to the decision-maker, and represents that decision-maker's rational credence about whether the individual has the unknown property Y. The decision is thus informed by two considerations: this credence, and the decision-maker's evaluation of the four possible outcomes—false positives, false negatives, true positives, and true negatives.[29] Given these evaluations, the optimal decision threshold will be the score at and above which it is more choiceworthy to take a certain action, below which more choiceworthy to take another. Decision-makers always face a tradeoff between false positives and false negatives, and the threshold that they set reflects how they evaluate this tradeoff. In this way, decision thresholds are value-laden.[30]

---

25 Dressel and Farid (2018). In fact, this paper finds that a predictive model that uses only age and number of prior arrests performs about as well as COMPAS.

26 The factors that COMPAS uses are not "race proxies" except in an weak sense of "proxy", by which any feature that correlates with race (e.g., income) would count as a race proxy.

27 For these other issues, the locus classicus is Thomson (1986). For recent discussion, see among others, Enoch, Spectre, and Fisher (2012); and Lippert-Rasmussen (2011).

28 When differential thresholds are called for, miscalibration can indirectly implement these differential thresholds—see the footnote above. That's why I don't claim that score calibration is a necessary condition in cases where differential thresholds are called for. In such cases, miscalibration might coincidentally implement the appropriate differential thresholds.

29 Again, this is an idealization for simplicity. While I identify the credences of decision-makers with the scores themselves, in fact decision-makers may of course consider other factors besides the scores.

The most austere way—by no means the only way—of modeling this tradeoff is in decision-theoretic terms. For example, suppose there are well-defined values (FP, FN, TN, TN) associated with each of the four outcomes, and that decision-makers ought to maximize expected value with each decision. Then the "expected value" of, for example, excluding or including a given strider with a given stride score would be given by:

$$EV_{exclude} = (p_{score})(TP) + (1-p_{score})(FP)$$
$$EV_{include} = (1-p_{score})(TN) + (p_{score})(FN)$$

Once we fix the values of TP, FP, TN, and FN, then there is some score threshold above which it is better in expectation to exclude, and below which it is better in expectation to include. From the two equations above, this threshold is given by:[31]

$$p *_{score} > \frac{TN - FP}{TN - FP + TP - FN}$$

This equation formalizes an intuitive principle: for any given assignment of values to the four outcomes, there is some probability p above which it is better in expectation to take a certain action, and below which it is better in expectation to take the other action.

We can weaken these very strong idealizing assumptions, of course. Indeed, the criminal justice context is especially complicated. First of all, needlessly detaining defendants not only harms them, but also arguably violates their rights against needless detention.[32] Thus, the decision involves both welfare and rights.[33] It is an open question how decisions under risk are supposed to handle such mixtures, and I do not settle this question here.[34] It will take a fully fledged moral theory to say how risks of harm and risks of rights violation should be, or indeed may be, balanced against each other.[35]

Secondly, the harms of errors fall on both defendants and non-defendants. Pretrial detention can harm the communities and families of defendants quite drastically. False negatives impose a risk on non-defendants, and also (via trial non-appearance) impede the functioning of the criminal justice system.[36] It is a substantive question how we should weigh the claims of defendants and the claims of non-defendants, and the extent to which the claims of different communities and different defendants may justify group-specific thresholds.[37]

---

30 See Kraemer, van Overveld, and Peterson (2011).

31 See Corbett-Davies et al. (ms) for a discussion of this result: "Thus, for a risk-neutral decision maker, the optimal action is [detain] when r(x) is sufficiently large, and otherwise the optimal action is [release]."

32 Incidentally, I think it's quite plausible that true positives are also rights violations –the practice of detaining defendants pretrial is hard to justify. In this paper, I have granted that it can be sometimes justified – because otherwise this decision problem is trivially easy: never detain!

33 Of course, many real-world applications of machine learning may not involve potential rights violations, and will be free of this complication.

34 As Hansson (2003) puts it, we can assign numerical values to the different outcomes, but these values would simply *report* what a fully fledged theory says; they would not themselves constitute a theory.

35 If one simply adds 'rights violations' into ones evaluation of outcomes and then does expected value, many might object to this kind of "utilitarianism of rights" as the wrong way of thinking of rights. For the classic treatment of this, see Nozick (1974).

36 Since these harms can compound over time, long-run effects must be taken into account.

Third, decision-makers may be risk-averse; or build in prioritarianism, desert, or some other values into the decision.

All that is to say, this simple expected-value framework is just one of many substantive positions about ethics under uncertainty. It is a non-trivial issue to say, in the COMPAS case and in many real-world cases, what exactly the right framework is, and thus what the right threshold will be. This is, indeed, one of many topics in ethics that become even more pressing in an age of algorithmic decision-making. As Lazar (2018) puts it, "In a world where advances in artificial intelligence have generated a compelling demand for formal versions of ethical theories adapted to decision-making under risk, it is more crucial than ever that deontologists explain how to systematically deal with duty under doubt."

Although I cannot settle these issues here, what's important for the topic at hand is that decision-makers must set some decision threshold(s), and that this threshold reflects how they evaluate the relative disvalue of false negatives and false positives. The worse false positives are, the higher one's decision threshold will be; the worse false negatives are, the lower. And if groups differ in how bad false positives or false negatives are for their members, differential thresholds may be appropriate; if they do not so differ, uniform decision thresholds are appropriate.

One central question in this paper is whether a decision procedure can be criticized as unfair in virtue of producing unequal false positive rates between groups with different base rates—and thus whether there is *pro tanto* reason to lower thresholds for groups with lower base rates. In the next section, I will argue that the answer to this question is "no".

But first, I'll show a sense in which calibration *is* necessary for fairness. To see how, consider a simple case where the harms of the four outcomes are relevantly similar for all individuals.[38] In this case, uniform decision thresholds are procedurally fair – they reflect that decision-makers are weighing the risks of false positives against the risks of false negatives in the same way for all individuals. Note what happens if scores are miscalibrated. For instance, suppose that COMPAS scores inflate the risk of rearrest for black defendants: a score of 8 is associated with a 80% chance of rearrest for white defendants, but a 60% chance for black defendants. Decision-makers, ignorant of this, uniformly detain all defendants with a score of 8 or more. In such a situation, a given black defendant at a score of 8 has a 40% chance of being a false positive, while a white defendant has a 20% chance of being a false positive. Thus, black defendants are subjected to a systematically higher risk of false positives.[39]

---

[37]There are substantive questions, beyond the scope of this paper, about which considerations we can take into account in justifying group-specific thresholds. One potential answer is that only defendant-affecting considerations can; it is inappropriate to take into account benefits to others. (May we be more willing to detain irritating people, who are missed less?) Another potential answer is that both defendant- and other-affecting considerations can. Yet another is that only *some* other-affecting considerations can be taken into account: namely, *other-affecting considerations specific to members of that group*. For this view, see Risse and Zeckhauser (2004).

[38]In fact, only the ratio $\frac{TN-FP}{TN-FP+TP-FN}$ must be equal; obviously there are other ways for this to happen than if all outcomes are equal.

[39]What about possible cases where miscalibration arises purely through chance? In any real-world context, sufficiently large numbers are at play that the likelihood that miscalibration is purely by chance becomes astronomically small, by Chebyshev's inequality. See Frick (2015) for a use of Chebyshev's inequality.

This procedure is effectively equivalent to a procedure which uses calibrated scores and an explicitly differential decision threshold: for example, a procedure where scores are calibrated, and decision-makers detain black defendants at 6 or above, and detain white defendants at scores of 8 or above.[40] This procedure would be a clear case of unjustified differential treatment, and a violation of procedural fairness. It treats black defendants' false positives as less serious than white defendants'.

Note that the unfairness of this sort of miscalibration—or, equivalently, of inappropriately differential thresholds—holds completely independently of group base rates. A predictor can selectively overpredict or underpredict for a group with any base rate whatsoever; a decision threshold can inappropriately weigh false positives and false negatives, for a group with any base rate whatsoever.

Of course, the costs of errors are not always the same for all defendants; in these cases, differential decision thresholds may be appropriate. If errors are differentially bad for members of different groups, then a group-specific differential decision threshold can be justified and even required. For instance, there are many reasons to believe that pre-trial detention is systematically worse for black defendants and black communities. This might justify differential decision thresholds.[41,42] But notice once again that this justification does *not* depend on the different base rates (and thus different false positive rates). It depends on the different costs of errors.

Finally, note that calibration is not alone sufficient for fairness. Not only can there be inappropriate thresholds, as mentioned, there can also be processes that are simply too inaccurate across the board. Even if the tool is calibrated, individuals may have a complaint that the procedure subjects them to higher risk of error than necessary. Or members of certain groups might argue that a particular (calibrated) procedure was selected in order to make them worse off than they would have otherwise been.[43] We can distinguish between the demand of *imposing as little risk as much as possible* (i.e. using as accurate a predictor as possible) and *imposing risks proportionally* (i.e. using a calibrated predictor that does not systematically impose more risk on certain groups of people).[44]

Still, we've seen why calibration is important to fairness: because miscalibration reliably leads to unjustified differential treatment.

### *4. Against false positive rate equality*

We've seen how group-differential interests can make group-differential decision thresholds appropriate. But equalizing false positive rates also mandates differential decision thresholds, independently of group interests: it mandates lowering the threshold for the group with the lower base rate, no matter their interests. For example, in the stride-height case it requires lowering the "exclude" threshold for women or raising the "exclude" threshold for men. This in turn would increase the number of false positives among women, while decreasing the number of false negatives; or increase the number of false negatives in men, while decreasing the number of false positives. If this adjustment shifts the threshold away from the group-interest-dependent threshold, then it will increase the total amount of disvalue from errors—by definition, since the interest-dependent threshold is the one that minimizes the expected disvalue from error.[45] This is, itself, a reason not to make this adjustment. But our question is: does fairness give us reason to make such threshold adjustments, in order to equalize false positive rates?

---

40 Both procedures, when faced with a white defendant and a black defendant who have the same chance of rearrested, will decline to detain the white defendant but will detain the black defendant.

41 See Corbett-Davies et al. (ms): "the costs of and benefits of decisions might vary across individuals, complicating the analysis. For example, it may, hypothetically, be more socially costly to detain black defendants than white defendants, potentially justifying group-specific decision thresholds". Huq (2019) makes related points.

42 As an added wrinkle, we could imagine a decision process that applied a uniform decision threshold, but on a miscalibrated scoring rule, and in so doing was equivalent to this appropriately differential decision procedure. But this would be a mere coincidence. In general, if the decision thresholds are appropriately set, miscalibration leads to worse decisions; when the decision thresholds reflect appropriate considerations of interests, then fairness requires calibration.

43 See Corbett-Davies et al. (ms) on redlining.

44 Compare to Broome's (1990) distinction between satisfying claims as much as possible, and satisfying claims proportionally.

I answer no. First, I note one relatively mundane reason that false positive equality might *seem* important to fairness, but is not: it simply sounds like calibration. Next, I give an argument for why false positive rates don't matter in themselves. Third, I rebut an argument that false positive rates track a kind of "equality of opportunity". Fourth, I canvass some important *indirect* ways that the false positive rate can matter—but not as a measure of fairness itself.

*4.1. The linguistic complication*

One relatively mundane reason that false positive rate inequality can seem important is simply that it *sounds* like miscalibration. Both measures can be expressed as "equal rates of people inaccurately predicted as high risk". This linguistic ambiguity can make these measures seem normatively similar. For example, Kleinberg et al. (2016) write that "the calibration condition and [false positive rate equality] seem to be asking for *variants of the same general goal*—that our probability estimates should have the same effectiveness regardless or group membership."[46] But once we realize that calibration and false positive rate equality are in fact directly opposed to each other, it no longer makes sense to think of them in this way. An argument must be given for thinking of false positive rates as a measure of procedural fairness.

*4.2. Complaints grounded in false positive rate inequality*

Consider, then, the following complaints from false positive rate inequality. They seem plausible on their face:

> "I am a male who is not TTFS, but I was excluded. False positive rate inequality shows that I was unfairly at more at risk of this false classification than a not-TTFS woman. After all, a greater share of not-TTFS men are false positives".
>
> "I am a black defendant who was not rearrested, but I was detained. False positive rate inequality shows that I was unfairly more at risk of this false classification than a non-rearrested white defendant. After all, a greater share of non-rearrested blacks are false positives."

However, these complaints go subtly wrong because they incorrectly link "risk of error" to the false positive rate. While miscalibration or inappropriately differential thresholds *are* evidence of systematically unequal risk of error, false positive rate inequality is not. To see why, it's helpful to once again fix the relevant statistical intuitions with a less charged example.

---

[45] As before, one might doubt whether the simple expected value framework is appropriate. The same caveat applies: even if it is not, there is a certain threshold that encodes the relationship between false positives and false negatives. And our question is whether, aside from these considerations, false positive rates give us a reason to adjust that threshold.
[46] Kleinberg et al. (2016, 3).

*Case: section grades*

Suppose that final papers could have a property of being a "true" A paper or a "true" B paper—the property of deserving an A or deserving a B. Suppose these are the only two grades in your course. Students in the course are divided into two discussion sections. Of course, often course sections happen to have different distributions of deserved grade. In this case, suppose that Section 2 has more “true B” papers.

Section 1: 10 true B papers, 20 true A papers
Section 2: 20 true B papers, 10 true A papers

You are a moderately fallible grader. When you apply your rubric and give a paper a B, this grade is correct 80% of the time: 20% of the times that you assign Bs, the paper which you gave a B was in fact a true A paper. Call these false positive errors *false Bs.*

We can thus say that the $p_{score}$ of your “B” grade is given by $p_{grade}$: 80%.

Let's assume that your rubric is calibrated: $p_{grade}$ is the same for both sections ($p_{grade1} = p_{grade2}$). Thus, consider what happens when you apply this rubric to two sections with unequal base rates of true Bs.

Section 1: You assign 10 Bs. 2 of these are false Bs. (since $p_{grade1}$ =80%.)
Section 2: You assign 20 Bs. 4 of these are false Bs. (since $p_{grade2}$ =80%.)

As we’ve seen, base rate differences mean that false positive rates will *necessarily* be higher for Section 2.

$$\text{FPR} = \frac{assigned\ B\ \&\ true\ As}{true\ As} = \frac{false\ positives}{true\ As}$$

$$\text{FPR section 1} = \frac{2\ false\ Bs}{\text{20 true A students}} = 10\%$$

$$\text{FPR section 2} = \frac{\text{4 false Bs}}{\text{10 true A students}} = 40\%$$

Suppose a false B from section 2—call him Filbert—comes and complains to you about your grading procedure: "I am a true A student. In section 2, 40% of true A students were given Bs. In section 1, only 10% of true A students were given Bs. Your grading procedure is procedurally unfair to members of Section 2. Because I was in section 2, I was four times as likely to get the wrong grade. I wish I had been in section 1."

I think it’s quite clear that this complaint cannot be correct, and that the inequality of false positive rates simply does not indicate any section-wise unfairness. While Filbert does have grounds to complain about the error, he does not have additional complaint that your procedure is unfair to Section 2 students in particular. Consider: does Filbert actually have reason, as he claims, to wish that he had been in section 1? No. By $p_{grade}$, Filbert would still have been subjected to a *20%* chance of being a false B. Now, if he had been in Section 1, he would have been one of 2 false Bs out of 20 true A students, instead of being one of 4 false Bs out of 10 true A students. But this makes no difference to his treatment and the risk he was subjected to. Filbert has absolutely no reason to prefer to be in section 1, given that the grading procedure is the same in either case.

To strengthen this thought, suppose that you had actually graded Section 1 papers slightly more harshly than Section 2; say, you assigned 12 Bs and your $p_{score}$ was 75%. In this case, the false positive rate would still be higher for section Section 2 (40%, to 15%). In this case, Filbert should actually *prefer* to be in section 2, even though its false positive rate is higher than Section 1's.

This intuition can ground an argument, which proceeds from the premise that Filbert has no reason to prefer to be in Section 1 over Section 2.

*No preference argument:*

1. *No preference:* When there is group-wise inequality of false positive rate, a higher false positive rate does not give members of a group any reason to prefer that they had belonged to a group with a lower false positive rate.
2. *No preference, no complaint:* If inequality of some metric Y does not give members of some group a reason to prefer that they belonged to another group, then members of this group do not have a procedural fairness complaint grounded in the inequality of metric Y.
3. *No complaint, no unfairness:* If no member of a group has a procedural fairness complaint grounded in the inequality of metric Y, then group-wise inequality of metric Y is not sufficient for procedural unfairness towards members of this group.

—
*Conclusion*: Group-wise inequality of false positive rate is not sufficient for group-wise procedural unfairness.

Thus, if one wants to defend the false positive rate as a measure of unfairness, one must explain how there can be section-wise unfairness even though Filbert lacks even a *pro tanto* reason to prefer to be in section 1. Also note that an argument of this form does *not* undercut the complaints of individuals in groups who have been subjected to unfair decision thresholds (whether via miscalibration or otherwise). Those individuals *can* rationally prefer that they have a different, more appropriate threshold applied to them.

Now we can consider the complaints above: suppose they were made by a man with a stride of 160; or a defendant with a COMPAS score of 8. Each of these scores is associated with an 80% chance—of tallness, or of rearrest. Each individual was subjected to a 20% chance of error. To be sure, these individuals can complain that *in fact* an error has been made. But they cannot argue that the false positive rate shows that they were in some sense subjected to a higher risk *in virtue of their group membership*. In each case, the risk of error is properly thought of as 20%—*not* as whatever the false positive rate is. (Just as Filbert's risk was properly thought of as 20%, not the false positive rate of section 2).

In sum: the accuracy of the predictor and the decision thresholds can be used to measure how much risk individuals are asked to bear, and can indicate when members of certain groups bear unjustifiably higher risks. In contrast, the false positive rate does not properly measure risk, and so does not measure anything important about procedural fairness.

### *4.3. The argument from implausible recommendations*

Returning to the case of section grades, notice that to equalize false positive rates would require selectively boosting the grades of papers in section 2 because of their higher base rate of Bs, or selectively lowering the grades of papers in section 1. Intuitively, there is not even a *pro tanto* fairness reason to do so. This suggests another argument:

1. If false positive rate inequality is a kind of procedural unfairness, then we always have some *pro tanto* fairness-based reasons to raise decision thresholds for the group with the higher base rate (or lower them for the group with the lower base rate).
2. We do not always have such reasons.
3. Inequality of false positive rate inequality is not a kind of procedural unfairness

Premise 1 is true because making these threshold changes just *is* what it means to equalize false positive rates. The justification for Premise 2 comes from reflection on cases where we quite obviously have no reason to make such changes—in fact, in many cases we have strong fairness-based reasons *not* to do so. For instance, it seems clear that the members of Section 2 should not have their grades raised, or the members of Section 1 have theirs lowered. Or consider a case where achieving false positive rate equality would further disadvantage the vulnerable group: the COMPAS + benefit case. In this case, achieving false positive rate equality would mandate making the decision threshold higher for black defendants—restricting the number of black defendants who are given the benefit, and making it so that white defendants are given the benefit when equally at-risk black defendants are not.

*4.4. False positive rate as equality of opportunity*

Finally, consider a common motivation for thinking that false positive rate equality is a measure of fairness. This motivation stems from a common intuition:

> Surely not-tall men and not-tall women should be at the same risk of false positives. And this is what false positive equality asks for.
>
> Surely not-rearrested black defendants and not-rearrested white defendants should be at the same risk of false positives. And this is what false positive equality asks for.

I think this is a misleading intuition. It gains its initial appeal from the fact that fairness can demand something resembling false positive rate equality, in related but importantly different cases. Imagine a case in which, unlike the cases at hand, true features are known and so there is no risk of error. Call these *certainty + lottery* cases.

> *Certainty + lottery case:* Suppose that there are 50 below-6-ft men and 100 below-6-ft women *and we know who they are*. Suppose that only 120 of these people can go on the trip; 30 must be excluded.

In this case, the fair decision procedure is an equal lottery: give every individual, man or woman, a 25% (30/150) chance of being excluded. If instead we had set up a lottery that was not equal, giving members of different social groups different chances of exclusion—for example, if men had a higher chance of being excluded—then this would be patently unjustified differential treatment. Notice that in this case, creating a fair lottery is the same as equalizing this ratio across groups:

> Fair lottery ratio: $\frac{excluded}{not\ TTFS}$

And notice that this number just is what the false positive equality rate measures—in cases of risk. That is, I think, why false positive rate equality has sometimes been called 'equality of opportunity' in the machine learning and fairness literature.[47]

But this certainty case simply is not the same kind of case as COMPAS case and the stride-height case. These are cases of risk, and the central impossibility arises only in cases of risk. In such cases, decision makers are not equalizing odds for people that they *know* are unnecessary-to-detain. Instead, they face a tradeoff with each decision between a false positive and a false negative, and must try to treat individuals fairly in light of their credences. In such cases, it doesn't make sense to ask decision-makers to directly equalize these odds. In the risk case, unlike the certainty case, they cannot equalize these odds except by applying differential thresholds, and increasing the total harm from errors.[48]

4.5. *When false positive rate can matter indirectly*

It should be noted that, although it is not a measure of fairness in the way that people have thought it is, there are ways that the false positive rate can be significant. First, false positive rate can track something important about community dynamics. If a community has a high false positive rate, this means that a high share of people in that community who do not need to be detained are being detained. This can, of course, have deleterious effects, perhaps compounding effects as the false positive rate rises. Thus, while we've been using a simplifying assumption that the expected value of each decision can be considered in isolation and then added together, in many contexts the badness of each individual false positive may depend on the total false positive rate. In such cases, this would mean that false positives are worse for groups that have a higher false positive rate; this greater harm would justify a higher threshold.

Secondly, as we have seen, unequal false positive rates can be evidence of miscalibration or of base rate differences (or both). Thus it will often be a symptom of underlying unfairness and inequality, which gives us reason to scrutinize the decision process.

Lastly, a high false positive rate might entrench discrimination if other actors are sensitive to the false positive rate itself, and make harmful generalizations based on it. This, too, could be a harm of false positives, and something to be taken into account in the threshold.

## 5. *Conclusion and implications*

I've argued that false positive rate inequality is not, in itself, a measure of unfairness. If it were, this would have far-reaching consequences: we'd always have pro tanto reason to selectively raise decision thresholds for groups with higher base rates, or selectively lower them for groups with lower base rates. But I've argued that setting thresholds at different levels for members of different social groups, according to base rates, makes decision thresholds lose track of the interests of individuals. For example, recall the COMPAS + benefit case. Achieving false positive rate equality in this case would mean selectively raising the decision threshold for giving the benefit to black defendants, and benefiting fewer of them. Cases like this make it especially vivid that taking positive rate equality *per se* to be desirable can recommend adjustments that are prima facie less fair, and that would harm disadvantaged groups.

---

[47] Hardt, Price, and Sebro (2016)

[48] Another problem for false positive rate equality as a notion of procedural fairness: it is not clear why it should be preferred to false negative rate equality, and achieving both at the same time is typically impossible. It's not clear why one should have priority over the other.

One reason that false positive rate equality has, thus far, seemed to be important is because of the particular features of the case in which it rose to prominence. One major complication of the COMPAS case is that there probably *is* miscalibration of a sort embedded in this process. As mentioned, the pre-trial decision is putatively made on the basis of the risk of *recidivism*, which COMPAS purports to measure. But recidivism is not the same thing as rearrest. There is quite plausibly a biased relationship between arrests and crime, due to racially biased patterns of policing. So, if decision-makers take themselves to have unproblematic access to the risk of *recidivism*, and not merely rearrest, then COMPAS scores are a miscalibrated measure. This miscalibration should be corrected.[49] But, as we've seen, this goal is different from adjusting the decision process in order to achieve false positive equality. (For example, rearrest measures can also inflate recidivism for groups with lower base rates than whites).

For all these reasons, in the context in which it rose to prominence, false positive rate equality mandates a plausible change to the decision procedure: selectively seeking to detain fewer black defendants. However, false positive equality makes this plausible recommendation for the wrong reasons: it happens to point in the "right" direction because the group with the higher base rate, and so a higher false positive rate, is also the relatively disadvantaged group, and the group subject to the various other decision-procedure biases. This is why positive rate equality fails so badly in other domains: because these features come apart. When they do, it requires group-specific adjustments that don't track the relevant features of the decision at hand. Thus, in many domains where false positive equality might be invoked on behalf of disadvantaged groups, it would require unmotivated and potentially harmful adjustments. This an especially critical point given the rapid expansion of automated prediction in society.[50]

For this reason, discussions of algorithmic fairness and bias will be better served by reserving the term "bias" for miscalibration or inappropriate thresholds. If "bias" is used also to refer to false positive rate inequality, then calling a decision process "biased" will carry no information: virtually every predictive processes will be "biased" in one way or the other. In combatting the unfairness that does exist in situations like pretrial detention, false positive rate equality will be a distraction, not a guide. And of course, none of these metrics should distract us from the following fact: while decision procedures may be implemented more or less fairly, many times the most fair thing to do is to abolish the procedure altogether. This is plausibly true of most pre-trial detention.

---

[49]For example, the COMPAS guide for users does not remind users that there is a non-trivial gap between what decision-makers want to know about (chance of recidivism), and what COMPAS can measure (chance of rearrest). Northpointe Practitioners Guide to COMPAS (2011).

[50]See Corbett-Davies et al. (ms) for a similar caution against taking false positive rate equality as a standard: "One must carefully design and measure the targets of prediction to avoid retrenching biases in the data. But, importantly, one cannot generally address these difficulties by requiring that algorithms satisfy popular mathematical formalizations of fairness." *Contra* these authors, however, I think that satisfying calibration on the appropriate measures, and ensuring adequate decision thresholds—which are popular formalizations—does help us avoid unfairness. Not all popular mathematical formalizations are inherently deficient.